\documentclass[a4paper]{article}
\usepackage{amssymb} 
\usepackage{amsmath} 
\usepackage{graphicx}   
\usepackage{amsthm}
\usepackage{wrapfig}
\usepackage{pgfplots}

\title{Multipliers: comparison of Fourier transformation based method and Synopsys design technique for up to $32$-bits inputs in regular and saturation arithmetics}
\author{Danila Gorodecky \\United Institute of Informatics Problems of NAS of Belarus \\ University of Bologna \\ email: danila.gorodecky@gmail.com}
\date{}
\begin{document}
\maketitle
\thispagestyle{empty}
\begin{abstract}
The technique for hardware multiplication based upon Fourier transformation has been introduced. The technique has the highest efficiency on multiplication units with up to 8 bit range. Each multiplication unit is realized on base of the minimized Boolean functions. Experimental data showed that this technique the multiplication process speed up to 20\% higher for $2-8$ bit range of input operands and up to 3\% higher for $8-32$ bit range of input operands than analogues designed by Synopsys technique.
\end{abstract}
\begin{section}{Introduction}
There is a variety of approaches to arithmetical operation of multiplication for hardware realization, but there is no universal approach for efficient hardware multiplication. The efficiency of each technique is limited by a number of conditions: bit ranges, number of multiplicands, area of implementation, special arithmetic (signed, unsigned, saturation arithmetic, residue number system, and etc.).

Our approach for designing $A\cdot B=R$ multiplier reminds of Lego constructing. Initially we develop a multiplication block (or blocks), and then build whole multiplier using these structural blocks, which we call monolithic multipliers.

Our technique is relevant to two known mathematical algorithms. First is called Karatsuba multiplication [1] and the second is Fourier transformation method (FTM) [2].

Karatsuba multiplication splits $A$ and $B$ into two vectors with the same length and then performs multiplication independently for each vector. But it works only for a big number of multiplication (for some hundreds bits numbers) [1].

A more similar technique is FTM. It is known that this approach is suitable for big numbers [2, 3] and implemented in arithmetic in modulo [3].

In this paper we focus on regular arithmetic and in saturation arithmetic multiplication. If in the regular arithmetic a result of multiplication $A\cdot B=R$ is $2n$-bits vector $R$, for $A$ and $B$ are n-bits vectors, then in saturation arithmetic a result has the same length as input operands. We consider multiplication from 2 to 32 bits of input operands.

This paper is organized as follows: in Section 2 we describe monolithic multipliers design technique and propose results of the synthesis; Section 3 dedicated to description of monolithic based multipliers in the regular arithmetic; Section 4 dedicated to description of monolithic based multipliers in the saturation arithmetic; in Section 5 we propose a technique of reducing of adders in hardware multiplication; Section 6 provides the results of the synthesis of monolithic based multipliers comparing with Synopsys analogues; the last section resumes our study.

All multipliers have been described on Verilog and synthesised (without place-and-routing) with Synopsys 2014 CAD on 28 nm technology in $Synopsys \ standard \ synthetic \ library$ with $compile \ ultra$ mode.  We propose no-memory technique and it is dedicated to unsigned multiplication.
\end{section}

\begin{section}{Design of Monolith Multipliers}
The name of monolithic multiplier refers to a holistic structure. An idea of monolithic multipliers synthesis concludes in the generating of small range multipliers from the truth table of Boolean functions. Initially we find out more suitable monolithic multiplier in small bit range. Afterward it is repeatedly implemented to design a final structure of multiplier.

It is a well known fact that linear growing of number on variables of a Boolean function lead to exponential growing of the truth table. Thus minimization of disjunctive normal form (DNF) for a function on more than 10-15 variables with a standard minimization is unacceptable. So we study monolithic multipliers with no more than 16 inputs and 16 outputs.

Realization of the multiplication is based on Boolean functions implementation, those we used some types of minimizations of Boolean functions: Espresso in the version 2.3 [4] and the ELS minimizer [5].

We used two options of Espresso minimization: \textit{exact} minimization and \textit{qm} (Quine–McCluskey) algorithm of minimization. The exact minimization is more powerful, but for 10 inputs and 10 outputs, i.e., for $A\cdot B=R$, where $A$ and $B$ are 5-bits vectors, minimization with exact option is not suitable. For a bigger bit range multipliers we used \textit{qm} option of minimization.

We implemented following ELS minimizer options: \textit{options, class, literals, Espresso with power consumption minimization}. In some cases \textit{class} and \textit{literals} minimizations showed more preferable results than Espresso.

Table 1 shows number of disjunctions in full DNF and in minimized DNF for all monolithic multiplier in regular and in saturation arithmetics.

\begin{table}[!htb]
\small 
    \caption{Comparison of the number of in full DNF and in minimized DNF in}
    \begin{minipage}{0.48\textwidth}
      \centering
\begin{tabular}[t]{|c|c|c|}
\multicolumn{3}{c}{a) regular arithmetic}\\
\hline
Multiplier         & Truth table & Minimized \\ \hline
$2\times 2\to4$            & 14          & 8         \\ \hline
$3\times 3\to6$  & 111         & 40        \\ \hline
$4\times 4\to8$  & 678         & 160       \\ \hline
$5\times 5\to10$ & 3733        & 629       \\ \hline
$6\times 6\to12$ & 18953       & 2435      \\ \hline
$7\times 7\to14$ & 92334       & 9194      \\ \hline
$8\times 8\to16$ & 434660      & 38957     \\ \hline
\end{tabular}
\end{minipage}
 \begin{minipage}{0.48\textwidth}
\centering
\begin{tabular}[t]{|c|c|c|}
\multicolumn{3}{c}{b) saturation arithmetic}\\
\hline
Multiplier         & Truth table & Minimized \\ \hline
$2\times 2\to2$  & 10          & 5         \\ \hline
$3\times 3\to3$  & 68         & 14        \\ \hline
$4\times 4\to4$  & 392         & 44       \\ \hline
$5\times 5\to5$ & 2064        & 143       \\ \hline
$6\times 6\to6$ & 10272       & 511      \\ \hline
$7\times 7\to7$ & 49216       & 1881      \\ \hline
$8\times 8\to8$ & 229504      & 6916     \\ \hline
\end{tabular}
\end{minipage}
\end{table}

We synthesized monolithic $2\times 2$, $3\times 3$, $4\times 4$, $5\times 5$, $6\times 6$, $7\times 7$, and $8\times 8$ multipliers and compared the results in the speed with Synopsys analogues. The comparisons of the speed of computations for monolithic multipliers and Synopsys are represented for regular arithmetic on Figure 1 and for saturation arithmetic on Figure 2.

\begin{figure}[h]
\caption{Speed of computations of monolithic multipliers and Synopsys in regular arithmetic}
\centering
\begin{tikzpicture}
\begin{axis}[
    width=380pt,
    height=120pt,
    x tick label style={/pgf/number format/1000 sep=},
    ymax=7,
   	ylabel={Frequency, GHz},
    xlabel={Type of multiplication},
    symbolic x coords={$2\times 2$, $3\times 3$, $4\times 4$, $5\times 5$, $6\times 6$, $7\times 7$, $8\times 8$},
    enlargelimits=0.1, 
    ybar=5pt,
    nodes near coords
]
\addplot coordinates {($2\times 2$,6.6) ($3\times 3$,4.16) ($4\times 4$,3.33) ($5\times 5$,2.5) ($6\times 6$,2.38) ($7\times 7$,2.12) ($8\times 8$,2.04)};
\addplot coordinates {($2\times 2$,6.25) ($3\times 3$,5) ($4\times 4$,3.85) ($5\times 5$,3.03) ($6\times 6$,2.5) ($7\times 7$,2.17) ($8\times 8$,1.75)};
\legend{Synopsys,Monolithic}
\end{axis}
\end{tikzpicture}
\end{figure}
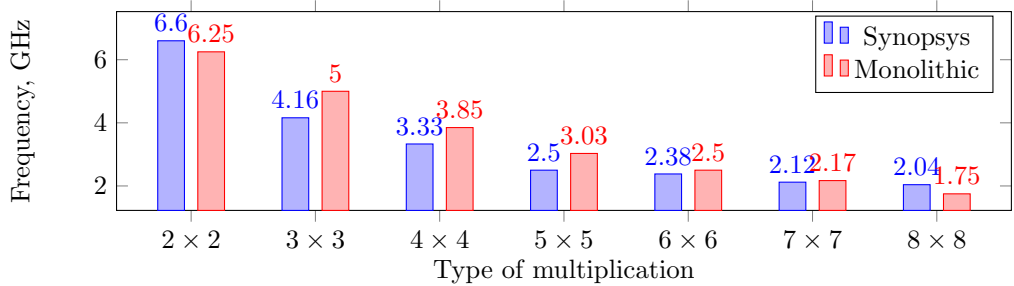

\begin{figure}[h]
\caption{Speed of computations of monolithic multipliers and Synopsys in saturation arithmetic}
\centering
\begin{tikzpicture}
\begin{axis}[
    width=380pt,
    height=120pt,
    x tick label style={/pgf/number format/1000 sep=},
    ymax=7,
   	ylabel={Frequency, GHz},
    xlabel={Type of multiplication},
    symbolic x coords={$2\times 2$, $3\times 3$, $4\times 4$, $5\times 5$, $6\times 6$, $7\times 7$, $8\times 8$},
    enlargelimits=0.1, 
    ybar=5pt,
    nodes near coords
]
\addplot coordinates {($2\times 2$,6.66) ($3\times 3$,5) ($4\times 4$,4) ($5\times 5$,3.12) ($6\times 6$,2.85) ($7\times 7$,2.7) ($8\times 8$,2.43)};
\addplot coordinates {($2\times 2$,6.25) ($3\times 3$,5.55) ($4\times 4$,5.3) ($5\times 5$,3.45) ($6\times 6$,2.78) ($7\times 7$,2.39) ($8\times 8$,2)};
\legend{Synopsys,Monolithic}
\end{axis}
\end{tikzpicture}
\end{figure}
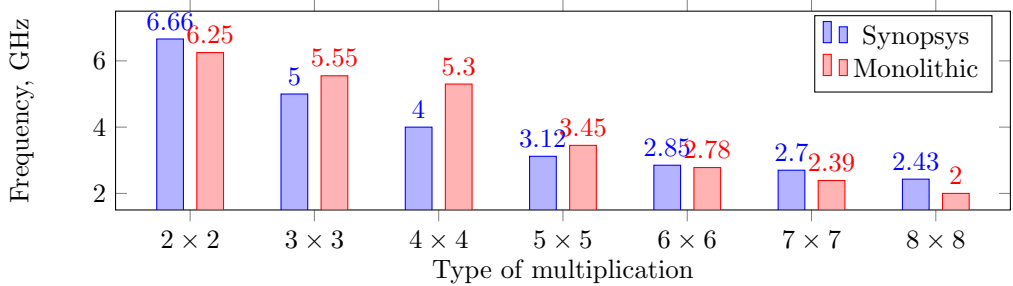

According to the comparison monolithic $3\times 3$ is faster on $20\%$, $4\times 4$ on 15\%, and $5\times 5$ on $21\%$ than Synopsys. As monolithic blocks we propose to use $4\times 4$ and $5\times 5$ multipliers, so we will use them to constructing blocks for monolithic based multipliers.
\end{section}

\begin{section}{Design of Monolithic Based Multipliers \\ in Regular Arithmetic}
An idea of the technique for design $A\cdot B=R$ multiplier reminds of Lego constructing. Initially we found out more suitable monolithic multiplier in a small bits range. Afterward we implement this monolithic block to $A\cdot B=R$ multiplier. It is know a similar approach for a regular multiplication which is a branch of Fourier transformation [2].

We represent arithmetical algorithm aimed to implement for hardware realization and based on FTM. The main idea of FTM consists in splitting the input vectors into $k$ groups with m bits in every group. So it is suitable for $k\cdot m$ dimension vectors. In common $A\cdot B=R$ can be represented as follows:
\begin{equation}\label{eq1}
A=\sum\limits_{i=1}^k {A_i\cdot 2^{m\cdot (i-1)}} \textrm{ and } B=\sum\limits_{j=1}^k {B_j\cdot 2^{m\cdot (j-1)}}, \textrm{ then } R=\sum\limits_{i=1}^k  \sum\limits_{j=1}^k {A_i\cdot B_j\cdot 2^{m\cdot (i+j-2)}}.
\end{equation}
According to Table 1 a) we take $m=3,4,$ and $5$.

Lets consider multiplication $A\cdot B=R$, where $A$,$B$ are 14-bits and $R$ is 28 bits. So we split each input $A=\left \{a_{14},a_{13},...,a_1\right \}$
and $B=\left \{b_{14},b_{13},...,b_1\right \}$ into three 4-bits and into one 2-bits vectors: $A_1=\left \{a_4,a_3,a_2,a_1\right \}, A_2=\left \{a_8,a_7,a_6,a_5\right \}, A_3=\left \{a_{12},a_{11},a_{10},a_9\right \}, A_4=\left \{a_{14},a_{13}\right \}, B_1=\left \{b_4,b_3,b_2,b_1\right \}, B_2=\left \{b_8,b_7,b_6,b_5\right \}, B_3=\left \{b_{12},b_{11},b_{10},b_9\right \}, B_4=\left \{b_{14},b_{13}\right \}$, where $a_{14}$ and $b_{14}$ are the most significant bits. Thus $A=\left \{A_4,A_3,A_2,A_1\right \}$ and $B=\left \{B_4,B_3,B_2,B_1\right \}$. Referring to the conditions of the example formula (\ref{eq1}) takes the following form:
\begin{equation}\label{eq2}
\begin{split}R=A_1\cdot B_1 + A_1\cdot B_2 \cdot 2^4 + A_1\cdot B_3 \cdot 2^8 + A_1\cdot B_4 \cdot 2^{12} + \\A_2\cdot B_1 \cdot 2^4 + A_2\cdot B_2 \cdot 2^8 + A_2\cdot B_3 \cdot 2^{12} + A_2\cdot B_4 \cdot 2^{16} + \\ A_3\cdot B_1 \cdot 2^8 + A_3\cdot B_2 \cdot 2^{12} + A_3\cdot B_3 \cdot 2^{16} + A_3\cdot B_4 \cdot 2^{20} + \\ A_4\cdot B_1 \cdot 2^{12} + A_4\cdot B_2 \cdot 2^{16} + A_4\cdot B_3 \cdot 2^{20} + A_4\cdot B_4 \cdot 2^{24}.
\end{split}
\end{equation}
According to (\ref{eq2}) the final phase of computation consists of fifteen operations of adding. It means that implementing in hardware the last step of computation will consists of a 4-level tree of fifteen adders.
\end{section}

\begin{section}{Design of Monolithic Based Multipliers \\ in Saturation Arithmetic}
The process of monolithic based multipliers in saturation arithmetic has a specific detail. In this type of arithmetic we are interested in the $n$ least significant bits of the result of multiplication, thus inputs and output vectors have the same length $n$.

We can use formula (\ref{eq1}) for multiplication in saturation arithmetic. It is clear that for $14$ by $14$ bits multiplication with 4-bits of splitting of inputs in saturation arithmetic we have the next redundant operands:$A_2\cdot B_4\cdot2^{16}$, $A_3\cdot B_3\cdot2^{16}$, $A_3\cdot B_4\cdot2^{20}$, $A_4\cdot B_2\cdot2^{16}$, $A_4\cdot B_3\cdot2^{20}$, $A_4\cdot B_4\cdot2^{24}$. But what about $A_1\cdot B_4\cdot2^{12}$, $A_2\cdot B_3\cdot2^{12}$, $A_3\cdot B_2\cdot2^{12}$, $A_4\cdot B_1\cdot2^{12}$? Results of these multiplications are 8-bits vectors, but we need only 4 least significant bits of them. In this case we use the multiplication modulo 4 and can represent a 14 by 14 multiplication as follows:
\begin{equation}\label{eq3}
\begin{split}R=A_1\cdot B_1 + A_1\cdot B_2 \cdot 2^4 + A_1\cdot B_3 \cdot 2^8 (mod\; 2^6) + A_1\cdot B_4 \cdot 2^{12}(mod\; 2^2) + \\+ A_2\cdot B_1 \cdot 2^4 + A_2\cdot B_2 \cdot 2^8 (mod\; 2^6) + A_2\cdot B_3 \cdot 2^{12} (mod\; 2^2) +\\ + A_3\cdot B_1 \cdot 2^8 (mod\; 2^6) + A_3\cdot B_2 \cdot 2^{12} (mod\; 2^2)+ \\ + A_4\cdot B_1 \cdot 2^{12} (mod\; 2^2).
\end{split}
\end{equation}
This manner of multiplication save $\frac{k^2-k}{2}-2$ adders comparing with (\ref{eq1}), where $k$ is a number of $m$-bits groups of subvectors of inputs. For example, in 14 by 14 bits multiplication with 4-bits splitting of the inputs we reduce number of adders from 15 to 9, and in 32 by 32 multiplication and the same splitting of inputs we save 26 adders.
\end{section}

\begin{section}{Adder-tree Levels Reduction Technique}
We proposed a technique to reduce the number of adding in the final step. The principle is to join in one vector as much as possible results of monolithic multiplications. According to the $14\times  14$ multiplication in regular arithmetic results of multiplications $A_1\cdot B_1=R_1$ and $A_1\cdot B_3\cdot 2^8=R_3\cdot 2^8$ will be represented as eight and sixteen bits vectors respectively, where $R_3\cdot 2^8$ includes eight zeros in the least significant bits. In this case $R_1$ and $R_3$ can be joint in one vector. Implementing this principle for (\ref{eq2}) we reduced number of adders from 15 to 6, and the adder-tree has been reduced from 4 to 3 levels.

Thus after replacement of $A_1\cdot B_1=R_1$, $A_1\cdot B_2=R_2$, $A_1\cdot B_3=R_3$, $A_1\cdot B_4=R_4$, $A_2\cdot B_1=R_5$, $A_2\cdot B_2=R_6$, $A_2\cdot B_3=R_7$, $A_2\cdot B_4=R_8$, $A_3\cdot B_1=R_9$, $A_3\cdot B_2=R_{10}$, $A_3\cdot B_3=R_{11}$, $A_3\cdot B_4=R_{12}$, $A_4\cdot B_1=R_{13}$, $A_4\cdot B_2=R_{14}$, $A_4\cdot B_3=R_{15}$, $A_4\cdot B_4=R_{16}$ and implementing of joining technique the result of multiplication will be represented with the next formula:
\begin{equation}\label{eq4}
\begin{split}R=(R_{16},R_{11},R_{3},R_{1}) + (R_{12},R_{7},R_{2},0000) + (R_{15},R_{10},R_{5},0000) + \\ + (R_{8},R_{6},00000000) + (R_{14},R_{9},00000000) + (R_{4} + R_{13},000000000000),
\end{split}
\end{equation}
where $R_1,R_2,R_3,R_5,R_6,R_7,R_9,R_{10},R_{11}$ are 8-bits vectors, $R_4,R_8,R_{12},R_{13},R_{14},R_{15}$ are 6-bits vectors, and $R_{16}$ is 4-bits vector.

The logic scheme of the adders tree of $A\cdot B=R$ multiplication, where A, B are 14 bits inputs and R is 28 output vector, is proposed on Figure 3.
\begin{figure}[h]
            \centering
            \includegraphics[width=1\textwidth]{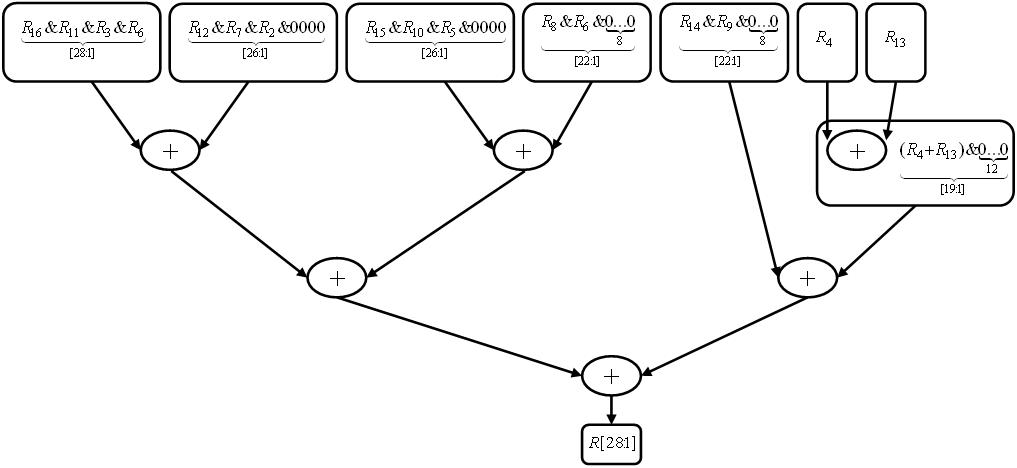}
\caption{Adders tree for 14 by 14 bits multiplication, where "\&" means concatenations, i.e. joining of binary vectors}
\end{figure}

In the common case an adders tree consists of $\rbrack\log_{2}M\lbrack$ levels, where $M=m\cdot k$. In Table 2 we compare number of adders in a common case and after final reduction for regular arithmetic a) and for saturation arithmetic b).
\begin{table}[!htb]
\footnotesize
\centering
    \caption{Comparison of the number of adders for the common case of multiplication and for the proposed technique in}
    \begin{minipage}{0.49\textwidth}
      \centering
\begin{tabular}[t]{|c|c|c|p{1cm}|}
\multicolumn{3}{c}{a) regular arithmetic}\\
\hline
Multipliers         & Common case & \begin{tabular}[c]{@{}c@{}}Reducing \\ technique\end{tabular} \\ \hline
$8\times 8\to16$  & 3         & 2         \\ \hline
$10\times 10\to20$  & 3       & 2        \\ \hline
$12\times 12\to24$  & 19      & 10       \\ \hline
$14\times 14\to28$ & 15      & 6       \\ \hline
$16\times 16\to32$ & 25      & 6      \\ \hline
$18\times 18\to36$ & 25      & 8      \\ \hline
$20\times 20\to40$ & 15      & 6     \\ \hline
$22\times 22\to44$  & 35      & 10         \\ \hline
$24\times 24\to48$  & 35      & 10        \\ \hline
$26\times 26\to52$  & 48      & 12       \\ \hline
$28\times 28\to56$ & 48      & 12      \\ \hline
$30\times 30\to60$ & 35      & 10      \\ \hline
$32\times 32\to64$ & 63      & 14      \\ \hline
\end{tabular}
\end{minipage}
 \begin{minipage}{0.49\textwidth}
\centering
\begin{tabular}[t]{|c|c|c|}
\multicolumn{3}{c}{b) saturation arithmetic}\\
\hline
Multipliers         & Common case & \begin{tabular}[c]{@{}c@{}}Reducing \\ technique\end{tabular} \\ \hline
$8\times 8\to8$  & 2         & 2         \\ \hline
$10\times 10\to10$  & 2       & 2        \\ \hline
$12\times 12\to12$  & 5      & 4       \\ \hline
$14\times 14\to14$ & 9      & 6       \\ \hline
$16\times 16\to16$ & 10      & 6      \\ \hline
$18\times 18\to18$ & 14      & 6      \\ \hline
$20\times 20\to20$ & 9      & 6     \\ \hline
$22\times 22\to22$  & 21      & 10         \\ \hline
$24\times 24\to24$  & 21      & 10        \\ \hline
$26\times 26\to26$  & 27      & 12       \\ \hline
$28\times 28\to28$ & 28      & 12      \\ \hline
$30\times 30\to30$ & 21      & 10      \\ \hline
$32\times 32\to32$ & 35      & 21      \\ \hline
\end{tabular}
\end{minipage}
\end{table}
\end{section}

\begin{section}{Results and Discussion}
We studied multiplication in regular and in saturation arithmetics: $8\times 8$, $10\times 10$, $12\times 12$, $14\times 14$, $16\times 16$, $18\times 18$, $20\times 20$, $22\times 22$, $24\times 24$, $26\times 26$, $28\times 28$, $30\times 30$, $32\times 32$, where $10\times 10$, $20\times 20$, and $30\times 30$ multiplications were realized with $5\times 5$ monolithic multipliers utilizing and the rest based on $4\times 4$ multiplication.

Figure 4 shows the results of the experiments in regular arithmetic.
\begin{figure}[h]
\caption{Synthesis of multipliers in regular arithmetic}
\centering
\begin{tikzpicture}
\begin{axis}[
    width=415pt,
    height=150pt,
    x tick label style={rotate=90,anchor=east},
    ymax=2.2,
   	ylabel={Frequency, GHz},
    xlabel={Type of multiplication},
    symbolic x coords={$8\times 8$, $10\times 10$, $12\times 12$, $14\times 14$, $16\times 16$, $18\times 18$, $20\times 20$, $22\times 22$, $24\times 24$, $26\times 26$, $28\times 28$, $30\times 30$, $32\times 32$},
    enlargelimits=0.05, 
    ybar=2pt,
    nodes near coords
]
\addplot coordinates {($8\times 8$,2.04) ($10\times 10$,1.85) ($12\times 12$,1.69) ($14\times 14$,1.66) ($16\times 16$,1.61) ($18\times 18$,1.53) ($20\times 20$,1.51) ($22\times 22$,1.38) ($24\times 24$,1.38) ($26\times 26$,1.38) ($28\times 28$,1.36) ($30\times 30$,1.31) ($32\times 32$,1.3)};
\addplot coordinates {($8\times 8$,2.08) ($10\times 10$,1.7) ($12\times 12$,1.7) ($14\times 14$,1.61) ($16\times 16$,1.61) ($18\times 18$,1.45) ($20\times 20$,1.39) ($22\times 22$,1.37) ($24\times 24$,1.35) ($26\times 26$,1.31) ($28\times 28$,1.28) ($30\times 30$,1.22) ($32\times 32$,1.24)};
\legend{Synopsys,Monolithic}
\end{axis}
\end{tikzpicture}
\end{figure}

The jittering of the frequency of calculation for proposed multipliers and Synopsys in the regular arithmetic is limited by: 8\% for $5\times 5$ based multipliers; 5\% for $4\times 4$ based  multipliers; 21\% for monolithic multipliers. The advantage of multipliers achieves: to 8\% by Synopsys comparing with the proposed; 1\% for $4\times 4$ monolithic based multipliers comparing with Synopsys; 21\% for monolithic multipliers comparing with Synopsys.

Figure 5 shows the result of the experiments in saturation arithmetic.
\begin{figure}[h]
\caption{Synthesis of multipliers in saturation arithmetic}
\centering
\begin{tikzpicture}
\begin{axis}[
    width=415pt,
    height=150pt,
    x tick label style={rotate=90,anchor=east},
    ymax=2.6,
   	ylabel={Frequency, GHz},
    xlabel={Type of multiplication},
    symbolic x coords={$8\times 8$, $10\times 10$, $12\times 12$, $14\times 14$, $16\times 16$, $18\times 18$, $20\times 20$, $22\times 22$, $24\times 24$, $26\times 26$, $28\times 28$, $30\times 30$, $32\times 32$},
    enlargelimits=0.05, 
    ybar=2pt,
    nodes near coords
]
\addplot coordinates {($8\times 8$,2.43) ($10\times 10$,2.17) ($12\times 12$,2.12) ($14\times 14$,1.96) ($16\times 16$,1.85) ($18\times 18$,1.75) ($20\times 20$,1.72) ($22\times 22$,1.66) ($24\times 24$,1.51) ($26\times 26$,1.53) ($28\times 28$,1.47) ($30\times 30$,1.45) ($32\times 32$,1.45)};
\addplot coordinates {($8\times 8$,2.5) ($10\times 10$,2.17) ($12\times 12$,2.12) ($14\times 14$,1.96) ($16\times 16$,1.81) ($18\times 18$,1.75) ($20\times 20$,1.61) ($22\times 22$,1.61) ($24\times 24$,1.56) ($26\times 26$,1.53) ($28\times 28$,1.47) ($30\times 30$,1.39) ($32\times 32$,1.43)};
\legend{Synopsys,Monolithic}
\end{axis}
\end{tikzpicture}
\end{figure}
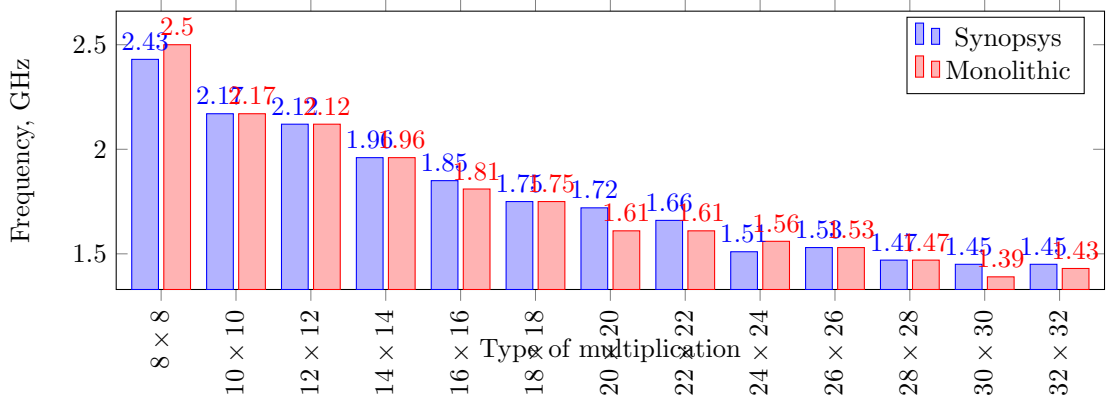

The jittering of the frequency of calculation for proposed multipliers and Synopsys in saturation arithmetic is limited by: 6\% for $5\times 5$ based multipliers; 3\% for $4\times 4$ based  multipliers; 33\% for monolithic multipliers. The advantage of multipliers achieves: to 4\% by Synopsys comparing with the proposed; 3\% for $4\times 4$ monolithic based multipliers comparing with Synopsys; 33\% for monolithic multipliers comparing with Synopsys.

The area of the proposed technique of monolithic based multipliers comparing with Synopsys jitters from 216\% (for monolithic based on $4\times 4$ multipliers) to 523\% (for monolithic based on $5\times 5$ multipliers) for both type of arithmetics.

Moreover we conducted experiments where Espresso multipliers have been changed by Synopsys $4\times 4$ and $5\times  5$ multiplication. In these cases area of the resulting multipliers oversized the Synopsys analogues in $5-10\%$, and in some cases even was smaller on $2-3\%$, but disadvantage in the speed was around $5-10\%$.
\end{section}

\begin{section}{Conclusions and Further Work}
We considered the approach of hardware multiplication. The approach concludes in the using of FTM relevant technique and in significant reducing of adders on the final step of multiplication.

The propose technique leads up to 20\% advantage in multiplication for monolithic blocks, and up to 3\% for multiplication from 8 to 32 input operands comparing with Synopsys analogues.

The results proposed in this paper are limited with $32\times 32\to64$ bit range. We have conjecture that implementation of the proposed multiplication technique will lead to more preferable difference for more wide bit ranges, i.e. $64\times 64$, $128\times 128$, and $256\times 256$.
\end{section}

\begin{section}{Acknowledgements}
The research was supported by the European Commission Erasmus Mundus MID.
The author is very grateful to Luca Benini and Micrel Lab for their help and support during the project. Thanks also to Paul Leonczyk and Petr Bibilo for cooperating in the conducted experiments.
\end{section}

\begin{section}*{References}
[1]{A. Karatsuba, Yu. Ofman: Multiplication of many-digital numbers by automatic computers, in Proceedings of the USSR Academy of Science, 1962, Vol. 145, No. 2, pp. 293-294, (in Russian).} \newline [2] {A. Schonhage, V. Strassen: Schnelle Multiplikation groser Zahlen, in: Computing 7 (1971), pp. 281-292.} \newline [3] {P. Gaudry, A. Kruppa, P. Zimmermann: A GMP-based Implementation of Schonhage-Strassen's Large Integer Multiplication Algorithm, in: Proceedings of the 2007 Interanational Symposium on Symbolic and Algebraic Computation (ISSAC'07), Waterloo, Ontario, Canada, pp.167-174, (2007).} \newline [4] {https://embedded.eecs.berkeley.edu/pubs/downloads/espresso/index.htm} \newline [5] {P.Bibilo, L.Cheremisinova, S.Kardash, N.Kirienko, V.Romanov, D.Cheremisinov: Automatizations of the logic synthesis of CMOS circuits with low power consumption: Programnaia ingeniria, 2013, Vol.8, pp. 35-41, (in Russian).}
\end{section}
\end{document}